\documentclass[preprint,showkeys, preprintnumbers,amssymb]{revtex4}

\usepackage{graphicx}
\graphicspath{{figs/}}
\begin{document}
\title{Minimum thermal conductance in graphene and boron nitride superlattice}
\author{Jin-Wu~Jiang}
    \altaffiliation{Electronic address: phyjj@nus.edu.sg}
    \affiliation{Department of Physics and Centre for Computational Science and Engineering,
             National University of Singapore, Singapore 117542, Republic of Singapore }
\author{Bing-Shen~Wang}
    \affiliation{State Key Laboratory of Semiconductor Superlattice and Microstructure and Institute of Semiconductor, Chinese Academy of Sciences, Beijing 100083, China}
\author{Jian-Sheng~Wang}
    \affiliation{Department of Physics and Centre for Computational Science and Engineering,
                 National University of Singapore, Singapore 117542, Republic of Singapore }

\date{\today}
\begin{abstract}
The minimum thermal conductance versus supercell size ($d_{s}$) is revealed in graphene and boron nitride superlattice with $d_{s}$ far below the phonon mean free path. The minimum value is reached at a constant ratio of $d_{s}/L\approx 5\%$, where $L$ is the total length of the superlattice; thus the minimum point of $d_{s}$ depends on $L$. The phenomenon is attributed to the localization property and the number of confined modes in the superlattice. With the increase of $d_{s}$, the localization of the confined mode is enhanced while the number of confined modes decreases, which directly results in the minimum thermal conductance.
\end{abstract}

\keywords{graphene, hexagonal boron nitride, superlattice, thermal conductance, confined phonon mode}
\maketitle

\pagebreak

Several recent experimental and theoretical groups have realized tunable electronic properties from different graphene and hexagonal boron nitride (h-BN) compounds.\cite{CiL,FanX,LiuY} With the further development of the synthesis technique, it will be quite possible for experimentalists to produce the graphene and h-BN superlattice (CBNSL). The superlattice has been proposed as a promising thermoelectric material (for review see eg. Ref.~\onlinecite{Cahill}). It can significantly manipulate the thermal transport property while only introduces neglectable effect on electron transport. As one of the most important features of the thermal transport in superlattice, the minimum thermal conductivity has been discussed by lots of researchers in the superlattice of various components.\cite{Venkatasubramanian,Simkin,Daly,Imamura,ChenY,Shiomi,Landry,Termentzidis} In all existing studies, the minimum thermal conductivity is achieved in superlattice with supercell size $d_{s}$ at about the phonon mean free path. In 2000, Simkin and Mahan explained the minimum thermal conductivity as the result of the competition between the particle and wave nature of the phonon modes in the superlattice.\cite{Simkin} The particle theory was applied for superlattice with $d_{s}$ larger than the phonon mean free path, while the wave theory was adopted in case of $d_{s}$ smaller than the phonon mean free path. These two different theories make opposite prediction on the behavior of thermal conductivity with the increase of $d_{s}$, and a combination of them results in a minimum thermal conductivity for $d_{s}$ in the order of phonon mean free path.

In this letter, we apply the nonequilibrium Green's function approach (NEGF) to investigate the thermal conductance versus $d_{s}$ in the CBNSLs, where $d_{s}$ is far below the phonon mean free path. We report a new minimum thermal conductance in the ballistic superlattice at $d_{s}/L\approx 5\%$, where $L$ is the total length of the superlattice. The value of the minimum point $d_{s}$ is proportional to the total length of the superlattice. It can not be explained by existing theory that predicts a constant minimum point $d_{s}$ around the phonon mean free path. We demonstrate that the minimum thermal conductance is related to various physical properties of the phonon confined mode (CM) in the superlattice.

\begin{figure}[htpb]
  \begin{center}
    \scalebox{1.1}[1.1]{\includegraphics[width=8cm]{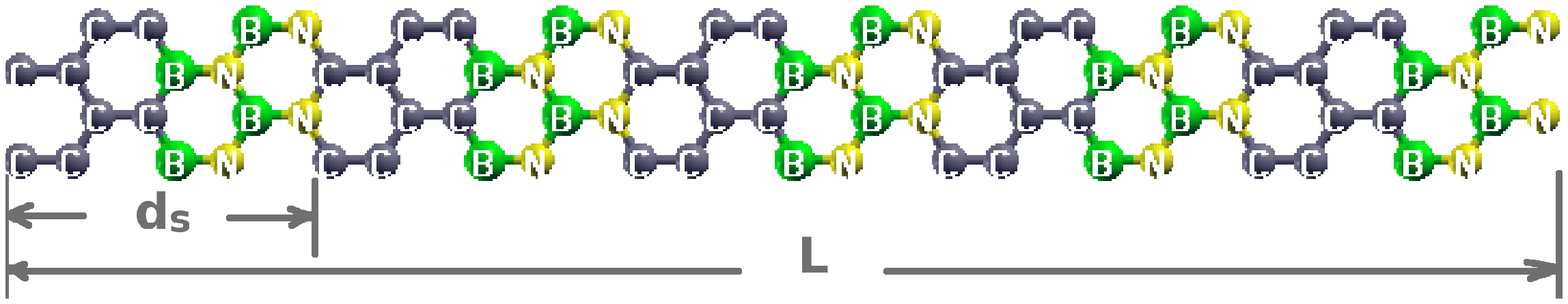}}
  \end{center}
  \caption{(Color online) The configuration of a CBNSL with total length $L=10$ and supercell length $d_{s}=2$ unit cells. Heat transport across the superlattice from left to right.}
  \label{fig_cfg}
\end{figure}

Figure.~\ref{fig_cfg} shows the configuration of a CBNSL. The size of the supercell is $d_{s}=2$ in units of the smallest translational period of the graphene nanoribbon ($d_{0}=4.26$~{\AA}). The total length of the superlattice is $L=10$. Due to large phonon mean free path,\cite{Ghosh,ChangCW} the phonon-phonon scattering can be ignored for phonon transport in each graphene and h-BN sheet segments of the CBNSL; thus it is in the ballistic transport region. We refer to this type of CBNSL as ballistic superlattice. The thermal transport in the CBNSL can be investigated by the ballistic NEGF approach.\cite{WangJS} The phonon thermal conductance is calculated by the Landauer formula: 
\begin{eqnarray}
\sigma_{\rm ph} & = & \frac{1}{2\pi}\int_{0}^{+\infty} d\omega \, \hbar\omega T[\omega]\left[\frac{\partial n(\omega,T)}{\partial T}\right],
\end{eqnarray}
where $T[\omega]$ is the transmission for phonon with frequency $\omega$, and $n(\omega,T)$ is the Bose-Einstein distribution function. The force constant matrix is obtained from the GULP\cite{Gale} with the Tersoff+UOOP potential.\cite{JiangJW1,JiangJW2} We have recently developed this set of potential for the graphene and h-BN sheet. It is efficient, stable and quite suitable for the study of heat transport.

\begin{figure}[htpb]
  \begin{center}
    \scalebox{0.75}[0.75]{\includegraphics[width=8cm]{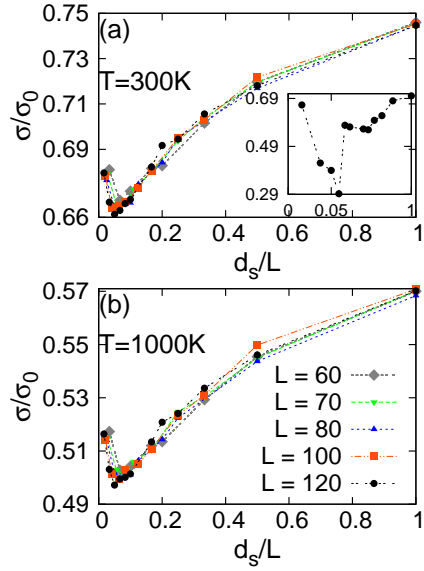}}
  \end{center}
  \caption{The thermal conductance versus supercell length $d_{s}$ at (a). 300 K and (b). 1000 K. $\sigma_{0}$ is the value of thermal conductance in pure graphene nanoribbon. Inset shows the results from molecular dynamics simulation at 300 K for $L=120$.}
  \label{fig_conductance}
\end{figure}

The thermal conductance in CBNSL versus $d_{s}$ is shown in Fig.~\ref{fig_conductance} at 300 K and 1000 K. The value of the thermal conductance is reduced by the thermal conductance in pure graphene nanoribbon with same size ($\sigma_{0}$). The conductivity $\kappa$ is related to the thermal conductance $\sigma$ through $\kappa=\sigma L /S$ with $S$ as the cross section area, so the ratio of the thermal conductivity $\kappa /\kappa_{0}$ will be exactly the same as $\sigma / \sigma_{0}$ shown here. Results of CBNSL with $L$ from 60 to 120 are shown in the figure. We have also observed similar phenomena in the CBNSL of different width and of zigzag edge configuration. A common feature in all curves is that the thermal conductance decreases with increasing $d_{s}$ in small $d_{s}$ region; then a minimum point is reached at $d_{s}/L\approx 5\%$; the thermal conductance increases with further increasing $d_{s}$ after the minimum point. The inset of panel (a) shows that molecular dynamics result also observes a minimum point around $d_{s}/L\approx 5\%$ for $L=120$. The value of minimum point $d_{s}$ increases with increasing $L$. For $L=120$, the minimum point is achieved at $d_{s}\approx 6$, i.e., 2.6 nm. This length scale is about two or three orders smaller than the phonon mean free path in the graphene and h-BN, which is around hundreds nanometers. Actually, a similar minimum thermal conductance was also observed in the ballistic isotopic graphene nanoribbon superlattice with $d_{s}$ around 2 nm.\cite{Ouyang} It is not the one predicted previously by Simkin and Mahan, which should be in superlattice with $d_{s}$ around the phonon mean free path.\cite{Simkin}

\begin{figure*}[htpb]
  \begin{center}
    \scalebox{0.7}[0.7]{\includegraphics[width=\textwidth]{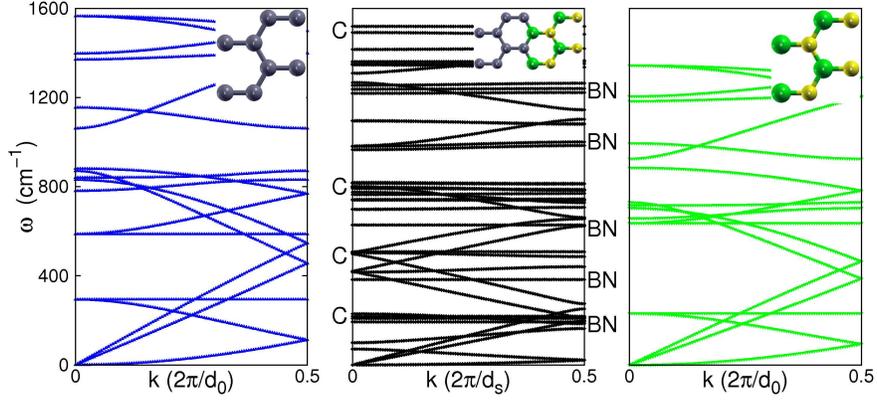}}
  \end{center}
  \caption{(Color online) Phonon spectrum in graphene nanoribbon (left), CBNSL (center), and h-BN nanoribbon (right). Insets are the corresponding unit cells. The symbols C/BN are with respect to the confined modes confined onto carbon/boron nitride periods.}
  \label{fig_dispersion}
\end{figure*}

Figure.~\ref{fig_dispersion} shows the phonon spectrum of the graphene nanoribbon (left panel), the CBNSL (center panel), and the h-BN nanoribbon (right panel). Inset in each panel displays unit cell of the structure. The abundant difference in the phonon spectrum between graphene nanoribbon and h-BN nanoribbon leads to lots of CMs in the CBNSL. The CMs can be confined onto either C period or BN period, which is indicated by C or BN in the center panel of the figure. For instance, the C-type CMs of highest frequencies are confined in C periods, corresponding to the in-plane optical phonon modes in graphene nanoribbon. In these C-type CMs, only C atoms vibrate while B and N atoms are silent. For the particular CBNSL in Fig.~\ref{fig_cfg}, the C-type CMs can be confined in each of the five C periods; thus there are five CMs with similar frequency. If the CMs are confined perfectly, i.e., all B and N atoms do not vibrate at all, then there will be no information exchange among these five CMs. As a result, they can not transfer heat current and have no contribution to the thermal conductance of the CBNSL. However, the confinement in real materials is not so ideal and the B, N atoms can join in slight vibration, which is much weaker than the vibration of C atoms. In this situation, the CM can exchange information with its neighboring CMs, which enables it to transfer heat energy from one C period to its neighboring C periods. This kind of non-perfect CMs have contribution to thermal conductance. As pointed out by Ghanbari and Fasol in 1989, the interaction between two CMs through the superlattice layer decreases rapidly with increasing layer thickness.\cite{Ghanbari} Hence, the interaction between two CMs becomes weaker  for larger $d_{s}$, which enhances the confinement of these modes. We refer to this mechanism as mechanism A. As a result of the mechanism A, the CMs make few contribution to thermal transport with increasing $d_{s}$. Another mechanism during the increase of $d_{s}$ with fixed $L$ is that the number of the C/BN interface will decrease, leading to smaller number of CMs. This mechanism is referred to as mechanism B. As a result of mechanism B, the thermal conductance increases with increasing $d_{s}$. The mechanism A is more important in the small $d_{s}$ region, which leads to the decrease of thermal conductance before the minimum point in Fig.~\ref{fig_conductance}. In the large $d_{s}$ region, the CMs are perfectly confined, so the mechanism B has dominant effect, leading to the increase of thermal conductance after the minimum point. The competition between these two mechanisms accounts for the minimum thermal conductance.

\begin{figure}[htpb]
  \begin{center}
    \scalebox{0.7}[0.7]{\includegraphics[width=8cm]{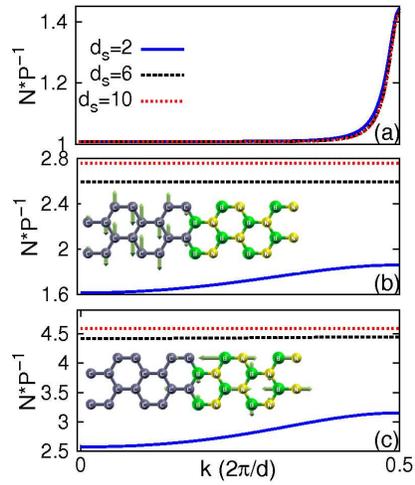}}
  \end{center}
  \caption{(Color online) IPR value for different phonon modes in CBNSL: (a). longitudinal acoustic mode, (b). C-type confined mode, and (c). BN-type confined mode.}
  \label{fig_ipr}
\end{figure}

To give a more quantitative characterization for the above argument, we investigate the localization properties of the CMs. The perfect CMs have better localization property than the non-perfect CMs. The localization property can be well described by the inverse participation ratio (IPR).\cite{BellRJ,JiangJW2010} The IPR for a phonon mode is defined through its normalized eigen vector $\textbf{u}$:
\begin{eqnarray}
P^{-1}=\sum_{i=1}^{N}\left(\sum_{\alpha=1}^{3} u_{i\alpha}^{2}\right)^{2},
\label{eq_ipr}
\end{eqnarray}
where $N$ is the total number of atoms, and $i=1,2,3$ are the three Cartesian coordinate components. The value of $P^{-1}$ can be determined straightforwardly for two ideal special cases. $P^{-1}=1/N$ for the three acoustic translational phonon modes, where all atom has the same vibrational displacement $u=1/\sqrt{N}$. $P^{-1}=1$ for phonon modes localized onto a single atom. A phonon mode with smaller IPR is better heat carrier. Fig.~\ref{fig_ipr}.~(a) shows the localization property of the longitudinal acoustic modes in CBNSL of different $d_{s}$. For convenience, we have plotted the $NP^{-1}$ instead of $P^{-1}$ in this figure. In whole Brillouin zone, the $NP^{-1}$ is about 1, indicating a good mobility of this acoustic mode. The acoustic modes are localized more strongly at the Brillouin zone edge. Considering the higher frequency of the acoustic modes at the Brillouin edge, this phenomenon is similar to acoustic properties of continuous media, where the evanescent decay rates are larger for modes of higher energy.\cite{Ghanbari} The localization property of the longitudinal acoustic modes is almost the same in superlattice with different $d_{s}$. We have checked that this is also true for other two acoustic modes. Panels (b) and (c) show the localization property of C-type and BN-type CMs. Insets display the vibration morphology of the CMs. We calculate the $NP^{-1}$ of these two CMs in case of various supercell length $d_{s}$. Different from the acoustic modes, the $NP^{-1}$ of CMs is sensitive to the value of $d_{s}$, and increases with increasing $d_{s}$. It manifests the enhancement of the localization property for the CMs in CBNSL with larger $d_{s}$. As a result, the CMs make fewer contribution to thermal conductance for larger $d_{s}$. This calculation demonstrates the origin of mechanism A.

\begin{figure}[htpb]
  \begin{center}
    \scalebox{1.0}[1.0]{\includegraphics[width=8cm]{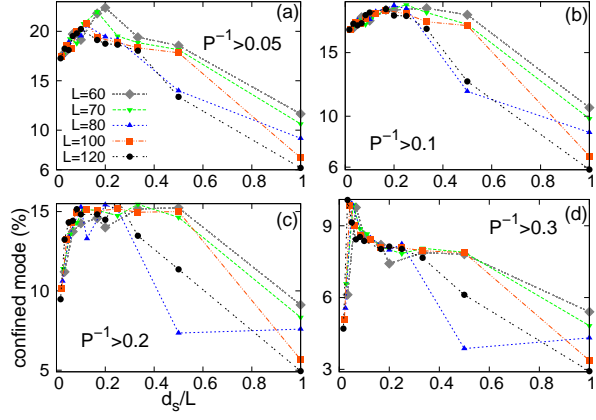}}
  \end{center}
  \caption{(Color online) The percentage of confined modes versus supercell length $d_{s}$ in CBNSL with different length. The confined mode is found by its IPR $P^{-1}>P^{-1}_{0}$, with $P^{-1}_{0}$ increases from 0.05 to 0.3 in panels (a)-(d).}
  \label{fig_percentage}
\end{figure}

As the perfect CMs could not transfer thermal energy, an even more straightforward route to explain the minimum thermal conductance in Fig.~\ref{fig_conductance} is to count the number of perfect CMs in the CBNSL of different supercell length $d_{s}$. Fig.~\ref{fig_percentage} shows the percentage of CMs in CBNSL of different length. A perfect CM is found if the IPR of this mode fulfills the criterion $P^{-1}>P^{-1}_{0}$. Four different values are chosen for $P^{-1}_{0}$: 0.05, 0.1, 0.2, and 0.3. Obviously, the percentage of CMs is smaller in case of larger $P^{-1}_{0}$. However, a common feature in all four panels is that the percentage of CMs has a maximum value. The above mechanisms A and B are both incorporated in this figure. On the left of the maximum point, mechanism A is important. With the decrease of $d_{s}$, some CMs turn to be able to pass heat energy to neighboring CMs, leading to smaller IPR value for these modes. As a result, the percentage of CMs decreases. On the right of the maximum point, the mechanism B is more important than mechanism A, since all CMs are already perfectly confined either onto C or BN period. With the increase of $d_{s}$ from the maximum point, the number of interface decreases, which leads to the decrease of the CMs percentage. The competition between mechanisms A and B gives rise to the maximum percentage of CMs. This maximum point explains the minimum thermal conductance in Fig.~\ref{fig_conductance}.

It should be noted, however, that the interface mode is another important type of localized mode, which exists in the SiGe superlattice.\cite{Ghanbari} The frequency of the interface mode locates in the middle of the energy gap between the phonon spectrum of the two components in the superlattice. A large value of energy gap is required for the existence of the interface mode. This condition is satisfied in the SiGe superlattice, as Si and Ge atoms have large mass difference leading to large energy gap. For the CBNSL studied in this work, the energy gap shown in Fig.~\ref{fig_dispersion} is not large enough to generate interface modes. Indeed, we do not see this type of modes in our calculation.

In conclusion, we perform a NEGF study of the thermal conductance in the ballistic CBNSL. We find a minimum thermal conductance in CBNSL with supercell size $d_{s}/L\approx 5\%$. The value of the minimum point $d_{s}$ is far below the phonon mean free path in graphene or h-BN. This minimum thermal conductance is not the one predicted by previous works where $d_{s}$ is about the phonon mean free path. We demonstrate that the CMs in the CBNSL play an important role in this phenomenon. On the one hand, the CM becomes more perfectly confined for larger $d_{s}$, leading to smaller thermal conductance. On the other hand, the number of the CM decreases with increasing $d_{s}$, resulting in larger thermal conductance. The former mechanism dominates the small $d_{s}$ region, while the later mechanism controls the large $d_{s}$ region. The interplay between these two mechanisms explains the ballistic minimum thermal conductance.

In this work, the $d_{s}$ of the minimum thermal conductance increases with increasing $L$, which reflects the ballistic nature of the phonon transport in the superlattice. The minimum thermal conductivity explained by Simkin and Mahan\cite{Simkin} occurs at $d_{s}$ around the phonon mean free path, which indicates the diffusive nature of phonon transport. An interesting open issue is to investigate the crossover from the ballistic minimum point to the diffusive minimum point.

\textbf{Acknowledgements} The work is supported in part by a URC grant of R-144-000-257-112 of National University of Singapore.


\end{document}